\documentclass[nohyperref]{article}

\usepackage{microtype}
\usepackage{graphicx}
\usepackage{subfigure}
\usepackage{booktabs} 
\usepackage{amsmath}
\usepackage{amssymb}
\usepackage{mathtools}
\usepackage{amsthm}
\usepackage{hyperref}
\usepackage{algorithmic}
\usepackage{algorithm}

\usepackage[capitalize,noabbrev]{cleveref}

\theoremstyle{plain}

\theoremstyle{definition}

\theoremstyle{remark}

\usepackage[disable,textsize=tiny]{todonotes}
\usepackage[textsize=tiny]{todonotes}

\begin{document}

\title{Deep Learning-based Spatially Explicit Emulation of an Agent-Based Simulator for Pandemic in a City}
\author{Varun Madhavan, Adway Mitra, Partha Pratim Chakrabarti}
\date{Indian Institute of Technology Kharagpur}

\maketitle

\vskip 0.3in

\begin{abstract}
Agent-Based Models are very useful for simulation of physical or social processes, such as the spreading of a pandemic in a city. Such models proceed by specifying the behavior of individuals (agents) and their interactions, and parameterizing the process of infection based on such interactions based on the geography and demography of the city. However, such models are computationally very expensive, and the complexity is often linear in the total number of agents. This seriously limits the usage of such models for simulations, which often have to be run hundreds of times for policy planning and even model parameter estimation. An alternative is to develop an emulator, a surrogate model that can predict the Agent-Based Simulator's output based on its initial conditions and parameters. In this paper, we discuss a Deep Learning model based on Dilated Convolutional Neural Network that can emulate such an agent based model with high accuracy. We show that use of this model instead of the original Agent-Based Model provides us major gains in the speed of simulations, allowing much quicker calibration to observations, and more extensive scenario analysis. The models we consider are spatially explicit, as the locations of the infected individuals are simulated instead of the gross counts. Another aspect of our emulation framework is its divide-and-conquer approach that divides the city into several small overlapping blocks and carries out the emulation in them parallelly, after which these results are merged together. This ensures that the same emulator can work for a city of any size, and also provides significant improvement of time complexity of the emulator, compared to the original simulator. 
\end{abstract}

\begin{figure}[ht]
\vskip 0.2in
\begin{center}
\centerline{\includegraphics[width=9cm, height=5cm]{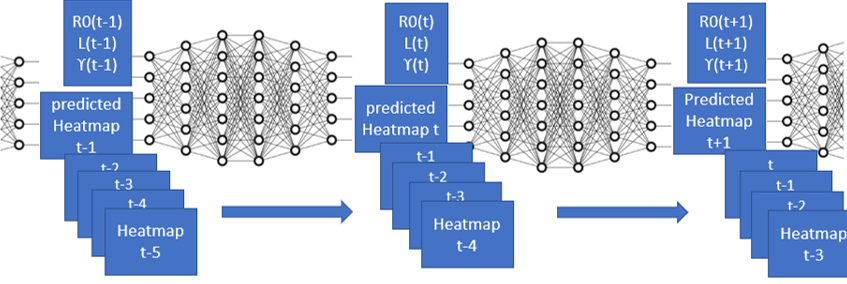}}
\caption{The scheme of emulation by producing heat-maps of daily new infections, hospitalizations, recoveries, deaths etc by a neural network based on past few days' heatmaps of the same, along with basic reproductive number $R_0$, lockdown policy $L$ and compliance rate $\gamma$ for each day}
\label{icml-historical}
\end{center}
\vskip -0.2in
\end{figure}

\section{Introduction}
During the Covid-19 pandemic throughout 2020-2021, Epidemiological Models gained considerable attention worldwide, as policymakers scrambled to predict the impacts of  Non-Pharmaceutical Interventions (NPIs) like lockdown orders, weekend curfews, containment zones etc. It is necessary to make a delicate trade-off between curbing disease spread and socio-economic disruption. To enable such a trade-off, it is necessary to have an estimate of the possible consequences of different policies, such as a \emph{what-if} analysis. Carrying out such an analysis is difficult because there is no closed-form analytical formula to express the possible results of a particular policy, at least not for the pandemic spread. The way a pandemic may spread through a city is related to its demography, geography, public spaces and the interaction between these. It becomes necessary to develop simulators that may represent how the infections can spread out over space and time, under different characteristics of the virus (basic reproductive number $R_0$), the policies in place (eg. lockdown in whole or parts of the city) and people's compliance with them. One way to build such simulators is by using \emph{differential equation-based compartmental models}, or by \emph{Agent-Based Models}. The former are simpler and more efficient to run, but are unable to incorporate individual-level behavior in response to NPIs and spatial nature of the infection spread, while these pros and cons are reversed for Agent-Based Models. 

In this paper, we look to develop machine learning-based emulators for Agent-Based Models to get the best of both worlds, being able to produce spatially explicit simulations taking individual behavior into account like Agent-Based Models, but in a significantly less time. For this purpose, we use a neural network based on Dilated Convolutional Neural Network that is trained on spatial heatmaps of daily infections, hospitalizations, recoveries, deaths as simulated by Agent-Based Models as a response to different parameters like $R_0$ and lockdown policies, and is able to produce such spatial heatmaps by sequential predictions starting from specified initial conditions. We show through detailed experiments that the spatial heatmaps produced by the emulator very closely resemble those generated by the Agent-Based Model, while requiring only a small fraction of time apart from a one-time training phase. For 14  simulation runs, the emulator (including training) takes equal time as the ABM, for 29 runs the emulator requires only half the time required by the ABM, and for 58 runs the time required by emulator is only a quarter of ABM. To the best of our knowledge, this is the first work that can successfully emulate a full-scale Agent-Based simulate at spatial scale. 

The rest of the paper is organized as follows - in section \ref{sec:related-works} we describe prior work on epidemiological modelling, Agent-Based Models and the emulation of ABMs. In section \ref{sec:problem-definition} we formally state the problem, followed by a description of working of the ABM used for epidemic modelling in section \ref{sec:Agent-Based-simulator}. In section \ref{sec:emulation-framework} we describe the main Emulation framework, and discuss experiments and results of this approach in section \ref{sec:experiments}. Finally, we discuss some of the possible applications of this framework in section \ref{sec:applications} and conclude in section \ref{sec:conclusion}. 

\section{Related Works}
\label{sec:related-works}

In this section, we review relevant literature on the three aspects of this work- Epidemiological models, Agent-Based Models and Machine Learning based surrogate modeling.

\subsection{Epidemiological Models}
Compartmental models have commonly been used for epidemic simulations, where the total number of individuals in different health states like Susceptible, Infected and Recovered are tracked, and the time-evolution of these numbers modeled by differential equations. In the context of Covid-19, these well-known compartments are insufficient due to the presence of large number of \emph{asymptomatic} cases. In these models, the impacts of NPIs is handled by varying the parameters (eg. rate of infection and recovery) over time. \cite{cov_1} proposes a COVID-ABS model to predict the impact of NPIs on public health and urban economics by simulating the contained environment of an urban city. The framework implements a compartmental SEIR model and takes into account the demographic and economic distributions of the population. Census data for Brazil is used for the parameters of this SEIR model, but the framework can be extended for any demographic. \cite{cov_3} perform a similar analysis for France. They use detailed attributes pertaining to the specific demographics and social contact structure in France and implement a statistical model for city spaces and transport. \cite{cov_4} came up with the SUTRA compartmental model which includes more compartments like asymptomatic cases.

\subsection{Agent-Based Modeling}

Agent-Based Modeling emerged as late as the 1990s in the domain of computational social science (\cite{abm_1, abm_2}) for simulation of artificial societies and their underlying interactions and outcomes. In such models, the behaviors of individuals (called agents) and their mutual interactions are represented mathematically, to observe the outcomes of systems which are difficult to express mathematically. Agent-Based Modeling has also been used in the domains of economics (\cite{abm_5}), ecology (\cite{abm_4}) and population biology (\cite{abm_6}). Agent-Based Modeling has tremendous potential for modelling urban systems, as detailed in \cite{abm_3} to simulate the interactions of all stakeholders with their urban environment. \cite{ca_urban} and \cite{ca_abm} explore temporal changes in urban sprawl and land-use compare ABMs with Cellular Automata-Based models. In these works, agents represent residents in a city and rule-based models are used to simulate their interactions \cite{ct_1,ct_2}. Recent work on ABMs focuses on modelling specific unique characteristics/aspects of cities in greater detail. Works such as \cite{plc_1, rs_1, tr_1, tr_2, tr_3, ct_3} have focused on modeling different aspects of urban life, including urban population growth, pollution and vulnerability to climate change, mobility and traffic management, resource planning and so on. Several ABM-based software packages like MATSIM \cite{MATSIM} and  UrbanSim \cite{URBANSIM} have also been developed. 

In the context of Covid-19 pandemic simulation, several works have attempted to study the spread of infections as a function of social interactions among individuals of a city. For such modeling of social interactions, various aspects of a city such as the family structures, residences, workplaces, spaces of social interactions etc can be represented at different levels of complexity or detail. Several works like \cite{cov_2, cov_8, cov_9} have developed such models for simulation of the pandemic
This work uses the ABM developed in \cite{our_paper} to compare the performance of the Neural Network-based emulators in predicting the outcomes of the counterfactual scenarios studied in the paper.

\subsection{Emulating Agent-Based Models}
Machine Learning and Deep Learning are increasingly being used for Agent-Based Modeling, primarily for parameter estimation using techniques like Approximate Bayesian Computation, or for developing surrogate models or emulators. This field is still quite nascent. \cite{zhang2021synergistic} and \cite{doi:10.1080/13873954.2021.1889609} explore how Agent-Based Models have been augmented using Machine Learning techniques. In \cite{CityMatrix} Machine Learning techniques are used to emulate the real-time flow of city traffic as simulated by an Agent-Based Model. They use a shallow convolutional neural network (CNN), which is able to predict the traffic flow significantly faster than the original ABM while maintaining comparable accuracy. \cite{surrogate_1} and \cite{surrogate_2} also propose using Neural Networks to emulate the outputs of an ABM epidemic model. \cite{surrogate_1} compares the performance of several candidate ML models as emulators and concludes that Neural Networks perform the best. 

None of the emulators mentioned above are spatially explicit, i.e. they produce only a time-series of different quantities, not their spatial maps. Also, a comprehensive study comparing time benefits and accuracy of emulation is lacking. The current work aims to address these gaps.

\section{Problem Definition}
\label{sec:problem-definition}

Let us formally define the problem that we aim to solve in this paper. We consider a city with $N$ residents. The city is divided into $K$ blocks, and they have populations $N_1,N_2,\dots,N_K$ such that $\sum_{k=1}^KN_k=N$. Each resident $i$ is provided with attributes such as age, family connections, health status and workplace, and mapped to a block $S_i$ as residence. A pandemic strikes the city which has basic reproductive number $R_0(t)$ on day $t$ (this index varies over time as the virus mutates and NPIs are put in place). At t=0, $I_0$ people are infected, who may belong to any block uniformly at random. Over the days, the infections spread, and on day $t$, $I_{kt}$ people in block $k$ are infected, $R_{kt}$ persons recover there, $D_{kt}$ persons pass away, etc. This number depends on the social interactions among the people and the NPIs imposed, like total or partial lockdown orders. However, an individual may follow orders with compliance rate $\gamma$. 

We have an Agent-Based Model $f$ which can simulate the multi-channel spatio-temporal sequence $X= \{X_{kti}\}_{k=1,t=1,i=1}^{K,T,L}$ as $X = f(N_0, R_0, \gamma)$. Index $k$ refers to a block, $t$ refers to a day and channel index $i$ refer to different compartments of the pandemic, such as new positive cases, recoveries and hospitalizations on each day in a particular block. Our aim is to develop a neural network $g$, which can predict $X_{kti}$ as $X_{kti} = g(\{X_{jt'l}\}_{j=1,t'=t-H,l=1}^{K,t-1,L},\{R_0(t')\}_{t'=1}^{t-1},\gamma)$. Here, $H$ is a time horizon or lookback window. Using this neural network, we aim to predict the full sequence $X$ of daily infections. In Section 4 we describe the Agent-Based Model $f$, and in Section 5 we discuss the emulator neural network $g$.

\section{Agent-Based Simulator}
\label{sec:Agent-Based-simulator}

In this section, we focus on the Agent-Based Model $f$ that was discussed above to generate the spatio-temporal sequence of daily infections. This Agent-Based Model is largely based on the work by~\cite{our_paper}. In this model, each individual in the city is considered as an \emph{agent}. Each agent has a set of attributes - his home, family, workplace etc. The agent interacts with other agents in the city and, if infected, spreads the virus to those he comes in contact with probabilities based on the duration and nature of the interaction. The Agent-Based Model has two main components: the city component and the infection component.

\subsection{City Component}
The city model has specific modules for the city structure, economic activities, transportation, education, healthcare. The city module specifies many kinds of spaces - residential, workplaces, marketplaces etc and also different kinds of facilities and services. The movements and interactions of individuals in these places is simulated. The Economic Module is sub-divided into sectors, which are further divided into sub-sectors. Each sub-sector consists of several Workplaces. Workplaces are locations where agents go for work. Workplaces are mapped to a location within the city, and each employed agent is assigned to a specific Workplace, where they travel regularly and interact with co-workers. Public transport facilities are also represented for simulating the movements of individuals and their interactions in the process. The healthcare module consists of a network of Covid-19 Hospitals, Covid-19 Healthcare centres, and Covid Isolation Centres. Each of these facilities is associated with a location and has a fixed capacity. The education module deals with schools and colleges, each of which is associated with a location and capacity. Each agent representing an individual who is a student is assigned to an educational institute (school or college), depending on age and residential area.

\subsection{Infection Component}
This component is based on \emph{health states} of individuals, and how these states can change due to infection by the virus. Each agent consists of a state belonging to a finite state machine, which is a vector of dimension 2. The first dimension represents the virus-related state i.e Healthy (H), Infected (IF) Recovered (R) and Dead (D). The IF state has two sub-parts namely Symptomatic (S) and Asymptomatic (A). We call them Virus-State (\(S_v\)). The second dimension represents mobility-related states which are Free (F), Out-of-City (O), Quarantined (Q), Isolated (I) and Hospitalized (HP). We call them Mobility-State (\(S_m\)) These two kinds of states are independent of each other, and the final state consists of a concatenated vector of these two.

Regarding the Virus-state, initially, every person is assumed to be healthy (H). When a person in state $H$ comes in contact with a person in state $IF$, the former gets infected with a certain probability. The viral load (VL) of a newly infected person is assumed to follow a Beta distribution, scaled appropriately. If this load is below a threshold, the person remains asymptomatic. Otherwise, the person undergoes a fixed incubation period after infection, after which they start showing symptoms. The peak infection period is sampled from a Gaussian distribution, after which the person may die or start recovering (depending on age and co-morbidity). The time taken to recover is also considered to follow a Gaussian distribution. Infection can spread from one person to another with a certain probability whenever they share the same physical space. This probability is a function of the contact duration and physical distance between them. This is why we have considered the number of work hours and physical gap as attributes of workplaces.

Additionally, we define Compliance Rate $\gamma$ as the proportion of the population actually following the policy in place. Higher compliance rates mean no two agents meet beyond their daily schedule (family, school, workplace). 

\subsection{Simulation Methodology}
The simulation proceeds by initializing the system, which includes setting up all the agents and their attributes, workplaces, medical facilities etc. This is done by sampling from probability distributions, such that all the attributes are consistent with each other. A small random subset $N_0$ of the population is infected with the disease as initialization. After that, we begin the simulation for the specified period. At each time step, we sample the movements of each agent according to a stochastic process based on their daily routine, whose distributions are based on the agent's workplace, residence and travel preferences. We track the interactions between pairs of individuals (when they come spatially close during their movements), and in such cases, the infection may take place stochastically as explained above. 

The Agent-Based Model produces a spatio-temporal sequence, producing the number of infections, hospitalizations, deaths etc in each block of the city on each day. We can represent the effect of spatial policies like block-wise lockdowns, containment zones, spatial distribution of resources, the effect of competing virus variants, etc. Using the spatial blocks as a grid structure, this spatio-temporal sequence may be visualized as a time-series of multi-channel heatmaps, with the pixel intensity value in each channel representing the value of a statistic (e.g. number of positive tested individuals, number of hospitalized individuals, etc.) or parameter (e.g. the population of the block. One grid heatmap is created after each day of simulation.

\section{Emulation Framework}
\label{sec:emulation-framework}

\begin{figure}[!ht]
  \centering
  \includegraphics[width=6cm, height=3cm]{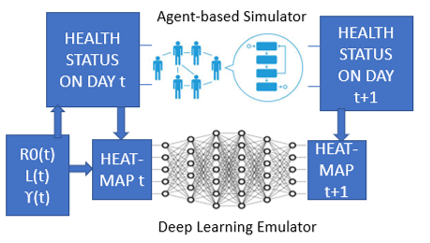}
  \caption{A block diagram showing how the deep learning-based emulator acts as a surrogate for the agent-based model}
  \label{figure:block}
\end{figure}

In this section, we discuss the Deep Learning based model to emulate the simulation results generated by the Agent-Based Model. The broad idea is to develop a Machine Learning model that will receive as inputs the initial number of infections (at $t=0$), the compliance rate $\gamma$, the Basic Reproductive Number $R_0(t)$ sequence, and the spatio-temporal sequence of the number of infections in each block of the city, over the past $W$ time-steps. The aim at each time-step is to predict the spatial heatmap of infections in the next time-step, acting as a surrogate of the ABM simulator as shown in Figure \ref{figure:block}. The individual-level details as used in the agent-based model are summarized as heat-maps for the deep learning-based emulator, along with inputs like $R_0, \gamma$ and lockdown information.

\subsection{Model for emulation}
\begin{figure}[!ht]
  \centering
  \includegraphics[width=7cm, height=3cm]{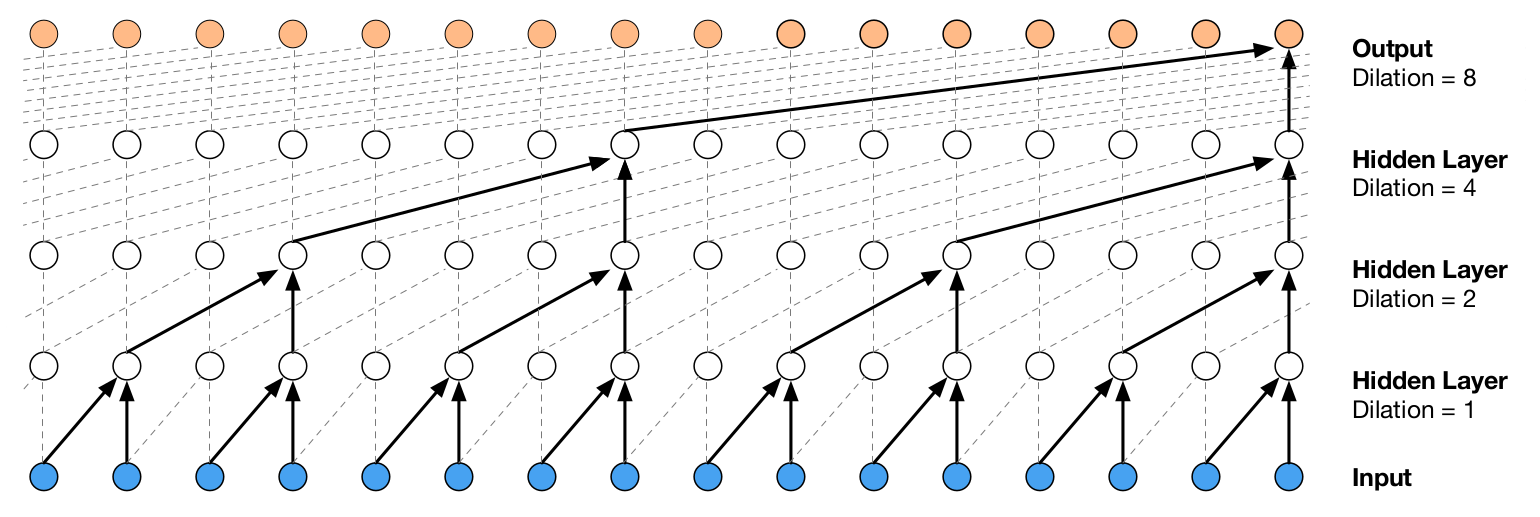}
  \caption{CNN with Dilations}
  \label{figure:cnn_with_dilations}
\end{figure}

For this purpose, we have tried out several network architectures that can leverage both spatial as well as temporal components in the heatmap sequences. A well-known Deep Learning model which can predict spatio-temporal sequences is the Convolutional LSTM~\cite{convlstm}. However, in our experimental trials it could not produce accurate heatmaps. After extensive trials with a variety of Deep Neural Network based models that use both recurrent and convolutional architectures, we find that a simplified version of the CNN with Dilations model presented in \cite{borovykh2018conditional}, modified to work with multi-channel heatmaps performs the best. We use \textit{Causal Padding} to ensure that the Network can only use past data to predict future values, and there is no leakage of data. Figure \ref{figure:cnn_with_dilations} shows the effect of using Dilations in a Convolutional layer.

It should be noted that the sequential predictions are made by feeding into the model previous predictions (heatmaps) by the model itself, not actual values. This is important because in the real-world use case we will need the emulator to make accurate predictions over a long period without having intermediate access to true values to correct previous mistakes.

\subsection{Divide-and-conquer Framework}
One of the key benefits of Deep Neural Networks is that modern GPU architectures allow for rapid inference by parallelizing matrix computations in batches. We use this ability to produce spatio-temporal trajectories for large cities using small amounts of training data. The Agent-Based simulator scales linearly as the number of blocks in the city. But instead of building emulators for different block sizes, we can use the same emulator by dividing the city into smaller overlapping regions, carrying out the emulation in each of them, and combining them by averaging. The emulations of these smaller regions can be done in parallel, and this gives massive time advantage. We call this as the \emph{Divide-and-Conquer} framework for emulation. As an example, we demonstrate how we can simulate a 20x20 \textit{Block} city using a 10x10 emulator in Algortithms 1 and 2. This procedure allows us to generate spatio-temporal trajectories for a 20x20 city using a 10x10 emulator that is significantly faster than an equivalent ABM, because we can generate all 10x10 predicted spatio-temporal heatmap sequences in parallel (rather than doing them one-by-one as we would have to do if we trained a 20x20 emulator instead).

\begin{algorithm}[tb]
\caption{Scaled ABM emulator Training}\label{alg:scaled-abm-training}
    \begin{algorithmic}[1]
        \STATE Generate training data using ABM simulations for a big city with blocks arranged as grids (size = (20, 20))
        \STATE Split this big city into regions (sub-grid size = (10, 10), stride = (2, 2))  
        \STATE Train a single emulator Model (block size = (10, 10)) to predict the spatio-temporal sequence of heatmaps of each region independently
    \end{algorithmic}
\end{algorithm}

\begin{algorithm}[tb]
\caption{Divide-and-Conquer ABM emulator Inference}\label{alg:scaled-abm-inference}
    \begin{algorithmic}[1]
        \STATE Use the Simulator to predict the case statistics for the first $H=5$ days
        \STATE Divide the initial city into regions (sub-grid size = (h, w) = (10, 10), stride = (sh, sw) = (2, 2))
        \STATE Use the learned emulator to predict the spatio-temporal heatmap sequence for the rest of the days for each region parallelly 
        \STATE Combine the predictions thus obtained for these regions to recreate the entire city's heatmap predictions. The prediction for each block on any day is the prediction for that day by all regions that overlap with that block
    \end{algorithmic}
\end{algorithm}

\section{Experimental Validation}
\label{sec:experiments}

Having described the models in detail, we now come to the experimental evaluation. The first step is to generate data by carrying out simulations using the Agent-Based Model. The next step is to use this data to train the deep learning-based emulator model, and test its performance.

\subsection{Data Generation}
The training data is generated by running the ABM repeatedly. Parameters like the virus \textit{Reproduction Number ($R_0$)}, population $N_0$, lockdown duration and compliance ration $\gamma$ are varied to generate distinct series. Since the ABM has a stochastic component, the simulation outputs for a particular parameter configuration vary slightly (about 5\% from the mean). To help the emulator learn the true signals and ignore the noise, we generate multiple series for each set of parameter configurations. 

We create a city with a population of $N=100000$ residents, distributed over a 10x10 gridded block structure. The population of each grid is sampled from a normal distribution with a mean of 1000 and a standard deviation of mean/6. To test the ability of the emulator to predict the case trajectory for different parameter values, we vary the $R_0$ between 1 and 4 in steps of 0.1. For each parameter value we generate several series of length $T=100$ days. All other parameters of the ABM are kept constant for this experiment. Of all the case-statistics tracked by the ABM, we retain 3 most descriptive ones - \textit{Cumulative Positive Tested}, \textit{Current Hospitalizations} and \textit{Current Asymptomatic Free}. Each heatmap is thus of size (10x10x3), and the length of the heatmap sequence is 100. The total number of such sequences generated for training is 512.

We split these sequences into training, validation and testing sets in the ratio 80:10:10 (410 + 51 + 51), uniformly across all parameter values. Each channel in $X$ is normalized by dividing it by the mean of the channel (mean calculated using only the training set). In order to train the emulator, we rearrange the training data as into (predictor, prediction) tuples, where each predictor is a sequence of $H$ consecutive heatmaps (10x10x3) and the parameters $(\gamma, R_0)$, and the prediction is the heatmap in the next timestep.  We set the lookback window as $H=5$. The loss function is the Mean Squared Error between predicted and actual heatmaps.

\subsection{Accuracy of Emulation}
\begin{figure}[!ht]
  \centering
  \includegraphics[width=0.4\textwidth]{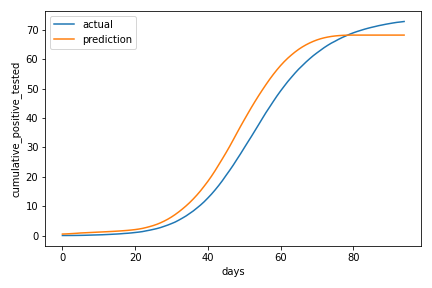}\includegraphics[width=0.4\textwidth]{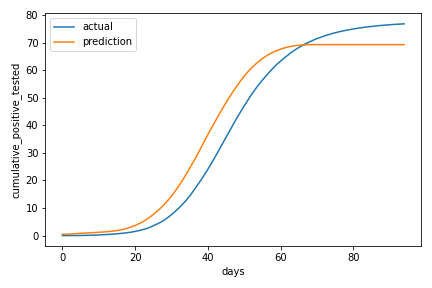}\\
  \includegraphics[width=0.4\textwidth]{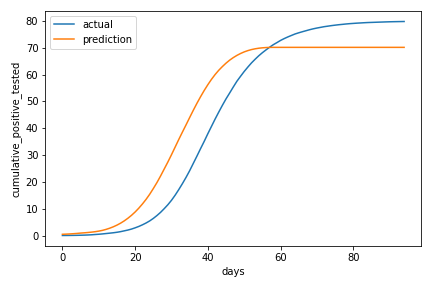}\includegraphics[width=0.4\textwidth]{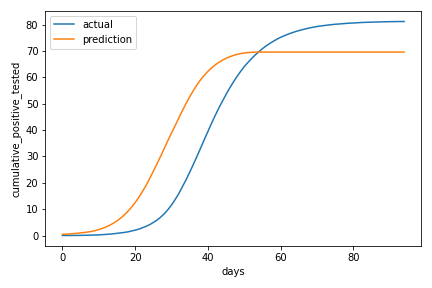}\\  
  \caption{Time-series of Cumulative Daily Infections for different $R_0$ values: clockwise from top-left: 1.8, 2.0, 2.2, 2.4. Blue: ABM, Red: emulator}
  \label{fig:temporal_baseline_1}
\end{figure}

\begin{figure}[!ht]
  \centering
  \includegraphics[width=0.45\textwidth]{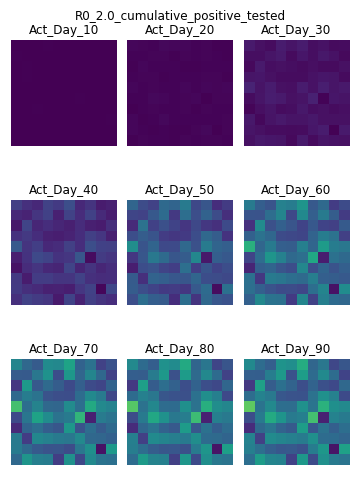}\hspace{0.5cm}\includegraphics[width=0.45\textwidth]{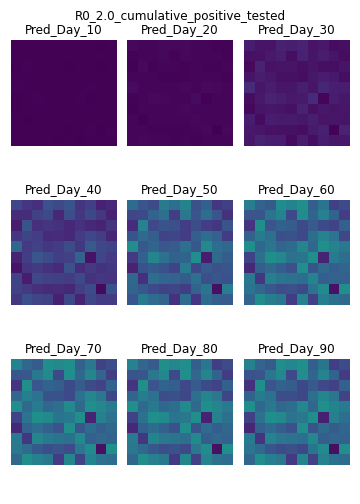} \\
  \includegraphics[width=0.5\textwidth]{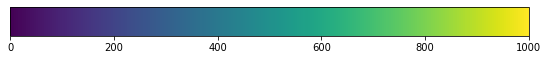}
  \caption{\textit{Cumulative Positive Tested} heatmaps of the city grid at 10-day intervals for $R_0=2$. Left: ABM (Act\_Day\_X), Right: emulator (Pred\_Day\_X)} 
  \vspace{-0.5cm}
  \label{fig:spatio-temporal-baseline-cpt}
\end{figure}

First of all, we compare the time-series of all the statistics as generated by the Agent-Based Model against those predicted by the emulator. Figure \ref{fig:temporal_baseline_1} shows the plot of the predicted (by emulator) values of the \textit{Cumulative True Cases} (red) compared to the ABM-simulated values (blue) for 4 random values of the \textit{Reproductive Number ($R_0$)}, and we can see the close similarity between the curves in all cases. We obtain similar plots for daily \textit{Hospitalizations} and \textit{Deaths} too (not shown for space constraints). Next, we also examine the spatio-temporal heatmap sequences generated by the ABM and compare them against the predictions by the emulator. In Figure~\ref{fig:spatio-temporal-baseline-cpt} we show the spatio-temporal heatmaps of the daily number of new infections detected in each block of the city, according to the emulator. These heatmaps closely match the heatmaps obtained from the simulator (not shown due to lack of space).

\subsection{Generalization}
We also wish to explore how the model performs when dealing with unseen parameter values.  Figure \ref{fig:temporal_interpolation_1} in the appendix shows the performance of emulating the ABM for unseen values of \textit{$R_0$} in-between the values it was trained on (the model was trained on values of \textit{$R_0$} between 1 and 4, in steps of 0.2). We observe that, as expected, it is able to perform well for these values of \textit{$R_0$}. On the other hand, Figure \ref{fig:temporal_extrapolation_1} in the appendix shows that the model is also able to emulate the ABM for unseen values of \textit{$R_0$} outside the range of $R_0$ values it was trained on. We observe that while it is able to perform well for larger values of \textit{$R_0$}, the emulator struggles for smaller values. This might be explained by the unstable behaviour of the ABM for smaller values of \textit{$R_0$}, since for very low \textit{$R_0$}, the infection may subside very quickly.
 

\subsection{Computational Benefits}
\begin{figure}[!ht]
  \centering
  \includegraphics[width=0.8\textwidth]{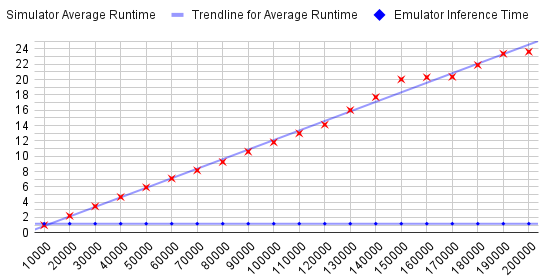}
  \caption{Time Analysis: Average Runtime vs. Population $N$ for simulator (red) and emulator (blue)}
  \vspace{-0.5cm}
  \label{fig:population-efficiency}
\end{figure}

The divide-and-conquer procedure allows us to generate spatio-temporal trajectories for a 20x20 city using a 10x10 emulator that is significantly faster than an equivalent ABM, because we can generate all 10x10 predicted spatio-temporal heatmap trajectories in parallel (rather than doing them one-by-one as we would have to do if we trained a 20x20 emulator instead). We show the time gain in Table 1, and the predicted spatio-temporal heatmap trajectory in the subsequent figures. Note that for $n$ simulation runs on a $20\times20$ city, the Agent-Based Model will require $38n$ seconds, while the emulation will require $504+0.76n$ seconds, which gives a major gain especially for high values of $n$. This is illustrated in Figure~\ref{fig:runtime} in the Appendix. Further, the Agent-Based Model also scales linearly in the number of agents, i.e. the city population $N$. In the next analysis, we vary $N$, keeping the grid size fixed at 10x10, and observe how the emulator compares with the simulator. Figure~\ref{fig:population-efficiency} shows that the emulator inference time remains nearly constant in all cases. 

\begin{table}[t]
\begin{center}
\begin{small}
\begin{sc}
    \caption{Computational performance of the simulator and the emulator in scaled inference. Both the simulation time and the emulator inference times are the times required to generate complete trajectories. All times are measured in seconds.}
    \begin{tabular}{|c|c|}
    \hline
            Task & \textbf{Simulator time} \\
            \hline
            \textbf{20x20 city-grid simulation time} & 38.0626 sec\\
            \hline\hline
            Task & \textbf{emulator}\\
            \hline
            \textbf{emulator training} & 503.93 sec \\
            \textbf{inference (10x10)} & 0.75 sec \\
            \textbf{scaled inference (20x20)} & 0.76 sec\\
            \hline
    \end{tabular}
    \vspace{-0.5cm}
\end{sc}
\end{small}
\end{center}
\end{table}


\section{Applications of Emulation}
\label{sec:applications}
The computational benefits obtained by the process of emulation may be used in situations where hundreds of simulation runs are required. Carrying out such runs using the computationally-expensive Agent-Based Model is highly time-consuming. Although training the emulator model takes time, it is a one-time process. The inference, which is very fast, needs to be carried out repeatedly which gives a major computational benefit compared to the Agent-Based Model. Below, we consider two situations where multiple simulation runs are needed, and demonstrate how the emulator scores over the simulator in such situations.

\subsection{Parameter Tuning}
In practical applications of epidemic ABMs, we often have historical observations of some statistics, e.g. the number of new cases, hospitalizations, recoveries, deaths, etc., but we do have any information about others, e.g. the virus $R_0$ or $\gamma$. Suitable values of these statistics are usually chosen by calibrating them based on the available historical data, using parameter search techniques like grid search, Bayesian Optimization, rejection sampling, etc. In this experiment, we demonstrate how the emulator can be used to speed-up this parameter search. 

We design an experiment where Bayesian Optimization is used to estimate the value of the virus $R_0$ from given historical data. A randomly sampled value of $R_0$ (uniformly sampled from the range [2.0, 3.0]) is used to produce one complete ABM simulation. Then, with this ground-truth value hidden, we use the emulator to line-search for this $R_0$ in a Bayesian Optimization paradigm. For each value of $R_0$ to be tested, a complete heatmap sequence is predicted by the emulator. The Mean Squared Error (MSE) of this predicted sequence vs. the actual heatmap sequence (i.e. the original ABM simulation) is used as the objective to be minimized by Bayesian Optimization. This process is then repeated using the ABM instead of the emulator to compare the time required.

This experiment is repeated for 20 distinct sampled values of $R_0$. Table \ref{table:parameter-tuning} compares the results when we use the emulator vs. the ABM (i.e. the simulator). We find that the $R_0$ searched by the ABM is slightly more accurate than by the emulator on average, but the latter provides a very significant time gain. Further, the emulator can make a better estimate of the $R_0$ in $40\%$ of the values tested. 

\begin{table}[t]
    \label{table:parameter-tuning}
    \begin{center}
    \begin{small}
    \begin{sc}
    \caption{Comparison of performances by Simulator and emulator to calibrate against observations by estimating $R_0$ value through Bayesian Optimization}
    \begin{tabular}{|c|c|c|}
    \hline
    & Average Error & Average Time Taken\\
    \hline
    Simulator & 15.82\% & 881 sec\\
    emulator & 25.27\% & 33.9 sec\\
    \hline
    \end{tabular}
    \vspace{-0.75cm}
    \end{sc}
    \end{small}
    \end{center}
\end{table}

\subsection{Scenario Analysis}
Another application in which a large number of simulation runs are usually needed is when we want to explore alternate intervention policies or counterfactual scenarios for \emph{what-if} analysis. In the context of a pandemic, we can consider the possible impacts of alternate non-pharmaceutical interventions like localized lock-downs and restricted use of public spaces, as discussed in \cite{our_paper}.

\begin{figure}[!ht]
    \centering
    \includegraphics[width=0.45\textwidth]{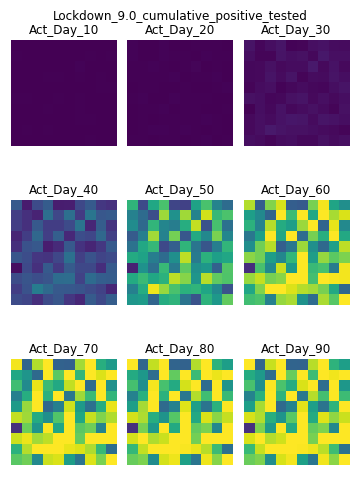}\hspace{0.5cm}\includegraphics[width=0.45\textwidth]{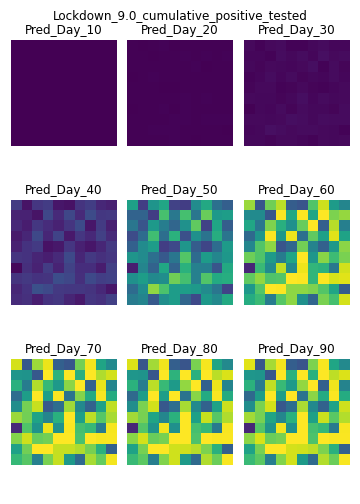}\\
    \includegraphics[width=0.5\textwidth]{colorbar.png}
    \caption{Scenario Analysis: Heatmaps for number of Cumulative Positive Tested persons, if there is a lockdown of 9 days from the 10th day. Left: observations from ABM simulator (actual values, Act\_Day\_X), Right: predictions by emulator (Pred\_Day\_X)}
    \vspace{-0.4cm}
    \label{fig:scenario-analysis-lockdown-9-cpt}
\end{figure}

In this experiment, we demonstrate that our emulator can learn the effect of lockdowns on the spatio-temporal case trajectory. For this experiment, the lockdown status of the city (i.e., whether the city is currently in a state of lockdown or not) at each time-step is passed to the emulator as an additional parameter channel. We consider lockdowns of various lengths in the range [3, 45], starting at a randomly chosen time-step near the start of the simulation (day 10), and observe the predictions of the emulator. We find that the emulator is still able to predict the spatio-temporal case trajectories accurately. In Figure~\ref{fig:scenario-analysis-lockdown-9-cpt}. It can be observed from the predicted trajectories that the emulator can learn the delay in rise of cases due to lockdowns quite accurately.

\begin{figure}[!ht]
    \centering
    \includegraphics[width=0.45\textwidth]{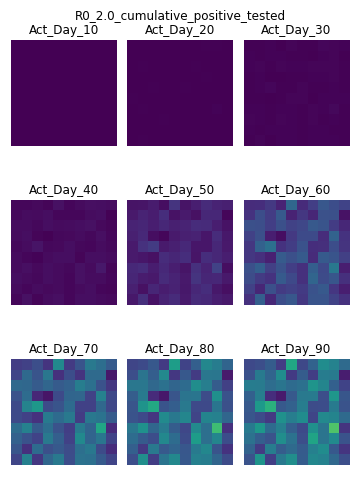}\hspace{0.5cm}\includegraphics[width=0.45\textwidth]{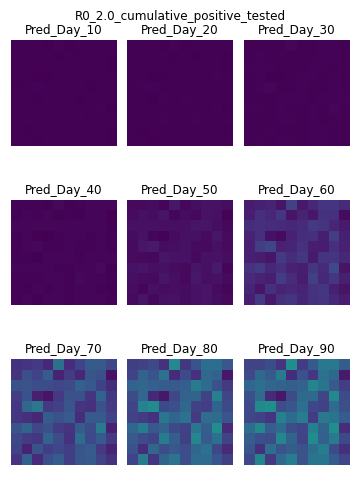}\\
    \includegraphics[width=0.5\textwidth]{colorbar.png}
    \caption{Effects of varying $R_0$ linearly from $1.0$ till $2.0$ and then back to $1.0$. Left: observations from ABM simulator (actual values, Act\_Day\_X), Right: predictions by emulator (Pred\_Day\_X)}
    \vspace{-0.1cm}
    \label{fig:scenario-analysis-varying-R0-2.0-cpt}
\end{figure}

The basic reproductive number $R_0$ does not stay constant during a pandemic. With increasing levels of immunity, either due to vaccination or infection, $R_0$ tends to decrease, while it may increase in the presence of multiple variants of the virus as it mutates. We want to see if the emulator can adapt to such changes in the $R_0$ time-series. Towards this end, we vary the $R_0$ through the simulation and observe the predictions of the emulator. The value of $R_0$ on each time-step of the simulation is passed as a feature to the emulator as another parameter channel. We demonstrate a simple trajectory for $R_0$ here as a proof-of-concept; it starts at a low initial value of 1, then linearly increases to a maximum value near the mid-point of the simulation, then again linearly decreases to the initial value. In addition to varying $R_0$ throughout the simulation, we also vary the maximum value (and hence the slope on either side) of the $R_0$ trajectory to observe how well the emulator can learn these effects. From the results shown in Figure~\ref{fig:scenario-analysis-varying-R0-2.0-cpt}, we observe that the emulator can predict accurate trajectories for a wide range of maximum R0s.

\section{Conclusion}
\label{sec:conclusion}

This paper introduced a deep learning-based emulator for Agent-Based Models for pandemic spread in a city. The proposed model can almost perfectly reproduce the Agent-Based Model's outputs in a spatially explicit way, taking into account dynamically varying parameters, at only a small fraction of time. This approach is particularly suitable when hundreds of simulation runs are required.

\nocite{langley00}

\bibliography{example_paper}
\bibliographystyle{ACM-Reference-Format}

\newpage
\appendix
\onecolumn
\section{Appendix}


\begin{figure}[!ht]
  \centering
  
  \includegraphics[width=6cm, height=4cm]{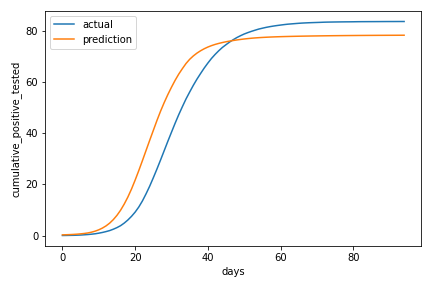}\includegraphics[width=6cm, height=4cm]{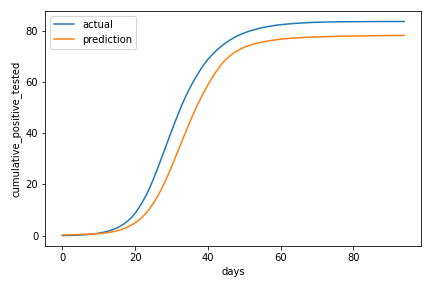} \\  
  \includegraphics[width=6cm, height=4cm]{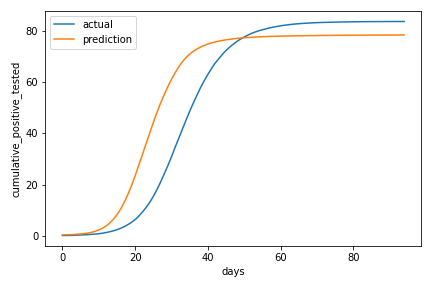}\includegraphics[width=6cm, height=4cm]{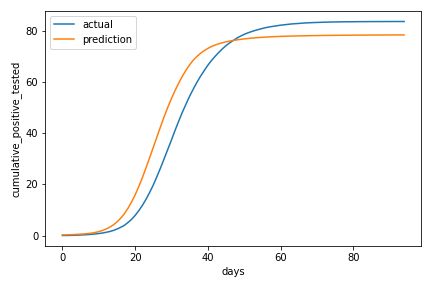} \\  
  \caption{Temporal Interpolation: Cumulative True Cases vs. Days for different unseen $R_0$ values inside the training range: clockwise from top-left: 2.65, 2.75, 2.85, 2.95}
  \label{fig:temporal_interpolation_1}
\end{figure}


\begin{figure}[!ht]
  \centering
  
  \includegraphics[width=6cm, height=4cm]{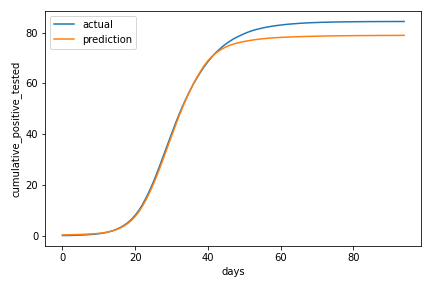}\includegraphics[width=6cm, height=4cm]{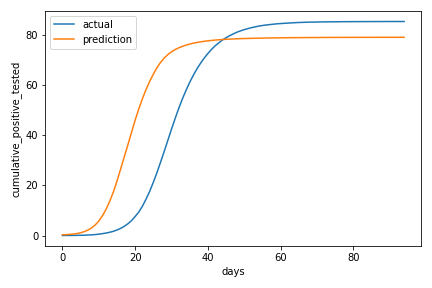} \\
  \includegraphics[width=6cm, height=4cm]{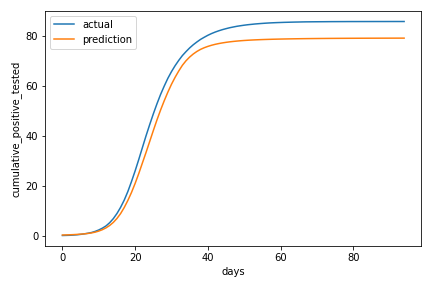}\includegraphics[width=6cm, height=4cm]{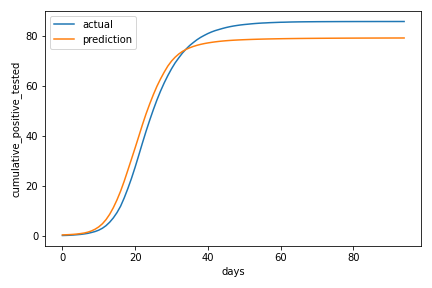} \\  
  \caption{Temporal Interpolation: Cumulative True Cases vs. Days for different unseen $R_0$ values outside the training range: clockwise from top-left: 3.05, 3.15, 3.25, 3.35}
  \label{fig:temporal_extrapolation_1}
\end{figure}

\begin{figure}[!ht]
  \centering
  \includegraphics[width=8cm, height=5cm]{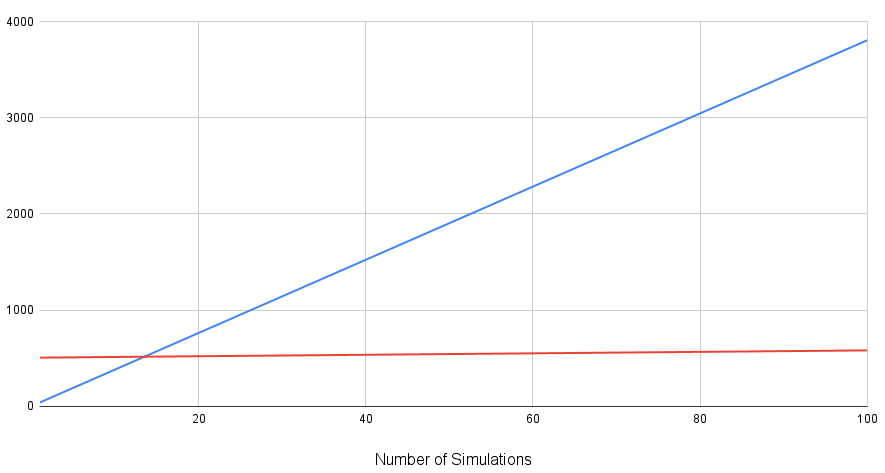}
  \caption{The time required in seconds (along Y-axis) for emulation (red), including the initial training time, increases negligibly with more simulation runs, and this makes it highly efficient compared to the agent-based simulator (blue)}
  \label{fig:runtime}
\end{figure}


\end{document}